\documentclass[a4paper,11pt]{article}
\pdfoutput=1
\usepackage{jcappub}
\usepackage[T1]{fontenc}
\usepackage[utf8]{inputenc}

\author[1,2]{Chris Clarkson,}
\author[2,3]{Roy~Maartens}

\affiliation[1]{Centre for Theoretical Physics \& Astronomy, Queen Mary University of London, London E1 4NS, United Kingdom}
\affiliation[2]{Department of Physics \& Astronomy, University of the Western Cape, Cape Town 7535, South Africa}
\affiliation[3]{National Institute for Theoretical \& Computational Sciences, Cape Town 7535, South Africa}

\emailAdd{chris.clarkson@qmul.ac.uk}
\emailAdd{rmaartens@uwc.ac.za}

\usepackage{amssymb}
\usepackage{amsmath}
\usepackage{bm}
\usepackage{mathtools}

\newcommand{\llangle}{\mathopen{\langle\!\langle}}
\newcommand{\rrangle}{\mathclose{\rangle\!\rangle}}
\usepackage{natbib,hyperref}

\def\up{\partial}
\def\lam{\lambda}

\newcommand{\ud}{\mathrm{d}}
\newcommand{\cd}{\mathrm{D}}

\def\si{\sigma}
\def\om{\omega}

\def\w{{\rm w}}

\def\dis{\displaystyle{\cdot}}
\def\pe{\perp}
\def\la{\langle}
\def\ra{\rangle}

\title{{Beyond dust in local cosmography: covariant kinetic theory}}

\abstract{Modelling cold dark matter as dust makes the matter congruence geodesic and
therefore removes any intrinsic dipole from the local Hubble flow.
The dust model breaks down as stream crossing develops, whereas a
collisionless gas  accommodates stream crossing via velocity dispersion in phase space.  We develop a covariant kinetic framework for this regime
without assuming a background spacetime or any geometrical symmetry.  Starting
from the exact massive Vlasov hierarchy, we construct its energy-integrated
moments
and organise these by velocity weight.  A self-consistent
truncation at the quadrupole
closes the dynamics on the matter energy density,
isotropic pressure and anisotropic pressure.  Their gradients generate a
matter-frame 4-acceleration, for which we find
an exact nonlinear solution
of the momentum constraint, valid in the general case without truncation.
This goes beyond previous work in which an arbitrary 4-acceleration is introduced.
Velocity
dispersion  also sources vorticity
through  the curl of 4-acceleration, and we derive the nonlinear expression for vorticity evolution.
We then propagate the non-geodesic matter flow into covariant
local cosmography, where the  4-acceleration is the intrinsic Hubble dipole and also enters
the  deceleration multipoles.  Our construction provides a covariant kinetic completion of
dust cosmography and identifies the quantities required to test collisionless
matter directly with local distance--redshift data. A companion paper
specialises the hierarchy to perturbed FLRW spacetime and predicts the size of
the resulting cosmographic signals.}

\begin{document}
\maketitle
\tableofcontents
\flushbottom
\allowdisplaybreaks
\[\]
\section{Introduction}
\label{s1}

Challenges to the standard model of cosmology, such as the Hubble tension, motivate the development of model-independent tests of the foundations of the standard model. In a series of papers \cite{Maartens:2023tib,Kalbouneh:2024szq,Kalbouneh:2024yjj,Kalbouneh:2025jnp}, we and others developed a covariant cosmographic framework for testing isotropy in the local Universe. The analysis is fully nonperturbative and does not prescribe a spacetime metric or background geometry, distinguishing it from most work, which assumes a  Friedmann-Lema\^itre-Robertson-Walker (FLRW) background.

In common with typical cosmographic analyses,   the matter in the local Universe is modelled in \cite{Maartens:2023tib,Kalbouneh:2024szq,Kalbouneh:2024yjj,Kalbouneh:2025jnp} as pressure-free dust. It was shown in \cite{Kalbouneh:2025jnp} that the covariant multipoles, which measure anisotropies in the expansion, could not be consistently fitted to the data at $z\lesssim 0.01$, corresponding to radial length scales $\lesssim 40\,$Mpc. This was interpreted as possibly a signal of the breakdown of the dust model on small scales.

In order to {develop a more realistic cosmography that allows us in particular to} test this interpretation, we need  a more general model of matter that recovers dust-like behaviour on large enough scales, but can probe below the scales where singularities arise for dust. The dust model assigns a unique velocity to an infinitesimal volume element at each spacetime event -- and this breaks down at stream crossing. However, gravitational collapse inevitably produces the crossing of particle trajectories (`shell crossing' in the case of spherical symmetry). Effectively this means multiple different velocities in  infinitesimal volume elements.

A natural kinetic continuation beyond single-stream dust is a collisionless gas of particles interacting only via the gravitational field that is generated collectively by all particles {(and by any other forms of energy that may be present)}. Multi-streaming, i.e. different particle streams in the same spacetime neighbourhood with different velocities, is accommodated by the inclusion of velocity dispersion in momentum space.
It is multi-streaming that generates isotropic and anisotropic stresses, as well as higher-order dynamical quantities, which are weighted averages over the kinetic distribution function. The pressure is not a collisional thermodynamic pressure; it is a kinetic stress generated by velocity dispersion.

The phase-space description remains regular through stream crossing and can accommodate the clustering and collapse into halos of cold dark matter. In practice, nonlinear collapse and the mean dispersion used in the companion perturbative analysis are calibrated with N-body simulations. At each order in the multi-streaming hierarchy, the evolution of a moment couples to neighbouring moments of the distribution function. Although moments beyond the quadrupole do not directly enter $T_{ab}$, and therefore the Einstein field equations, they affect the evolution of the lower moments in $T_{ab}$ via the collisionless Boltzmann, or Vlasov, equation.

Most of these features of the self-gravitating collisionless gas have been identified and investigated in the framework of perturbed  FLRW spacetime (e.g. \cite{Ma:1995ey,Buchert:1997dr,Buchert:1999dv,Valageas:2001df,Pueblas:2008uv,Erschfeld:2018zqg,Erschfeld:2020blf,Garny:2022tlk,Garny:2022kbk,Erschfeld:2023aqr,Nascimento:2024hle,Sipp:2025osk}). Related to this are phenomenological effective fluid descriptions (e.g. \cite{McDonald:2009hs,Peebles:2009hw}), including those based on effective field theory of large-scale structure (e.g. \cite{Baumann:2010tm,Carrasco:2012cv,Nascimento:2024rbv,Ivanov:2026nlf}).
{The two procedures are quite different: the kinetic stresses considered here are obtained by taking momentum moments of the phase-space distribution at each spacetime event, whereas an effective fluid description is obtained by coarse-graining over spatial scales and integrating out short-wavelength modes. In the kinetic theory approach, the pressure and anisotropic stress are components of the local energy-momentum tensor at each
point, irrespective of any spatial coarse-graining that is later introduced. By contrast, in effective field theory, isotropic pressure
and anisotropic pressure are \emph{effective} terms generated by integrating out short spatial modes, which enter the long-wavelength matter and gravitational equations in the same way as physical pressure.} This effective approach also assumes an FLRW background.
The effective coefficients encode short-scale dynamics and require matching,
often to simulations.
(A different approach to beyond-dust models, based on `matter horizons', is developed in \cite{Umeh:2026ajv,Umeh:2026mac,Umeh:2026cbn}.)

Here we focus on improving the covariant cosmographic analysis of the local Universe, which does not assume a background spacetime and is nonperturbative. Previous work has assumed a dust model. In \cite{Kalbouneh:2025jnp}, it was found that the covariant cosmographic parameters provided a well-behaved fit to Cosmicflows-4 data for $0.01<z<0.1$.  The fit to data started to break down for $z<0.01$, possibly indicating the presence of velocity-dispersion effects beyond the dust model. In order to more accurately interpret measurements
on scales where the dust model begins to break down, the matter model needs to include velocity dispersion.

The ingredients of this construction have previously appeared separately.
Covariant kinetic hierarchies provide the dynamics of collisionless matter
\cite{1983AnPhy.150..455E,ETM1983,Maartens:1999mf,Lewis:2002nc}, while general-spacetime
cosmography identifies
Hubble and distance--redshift
multipoles \cite{Maartens:2023tib,Kalbouneh:2024szq,Kalbouneh:2024yjj,Kalbouneh:2025jnp,Kristian:1966,MacCallum:1970,Clarkson:1999zq,Clarkson:2010uz,Umeh:2013,Heinesen:2020bej,Heinesen:2021azp,Heinesen:2023lig,Macpherson:2025qec,Pasten:2026rgb}.
The new step here is to connect these
structures dynamically.  We identify the cosmographic matter frame with the
energy frame of a massive collisionless distribution, derive its 4-acceleration
from the kinetic stresses, and close the nonlinear energy-integrated hierarchy
consistently through second velocity order.  This gives a physical kinetic
origin for the intrinsic Hubble dipole and for the stress corrections to the
deceleration multipoles, without introducing an FLRW background.

The exact  energy-integrated hierarchies are developed first, before the
second-moment closure is imposed.  We then derive the pressure, 4-acceleration
and vorticity equations and connect them to covariant Hubble and deceleration
parameters.  The companion paper \cite{Clarkson:2026xxx}
takes the linear scalar limit
about FLRW, supplies the mean dispersion from nonlinear halo formation, and
quantifies the resulting signals.\\

{\em Conventions:} Tensor indices $a,b,\dots=0,1,2,3$ indicate arbitrary spacetime coordinates or tetrad labels. We use signature $-+++$ and natural units, $c=1$.
{For late-time cosmography we neglect radiation, assume that baryons are cold and that dark energy is the cosmological constant. We only use the Einstein equations in \autoref{s5}.}

\section{Covariant kinetic theory of a collisionless gas}
\label{s2}

We employ a covariant form of relativistic kinetic theory, which is fully nonperturbative and does not assume a background geometry or any symmetries in the spacetime geometry. This approach was presented in the foundational papers by Ellis, Treciokas and Matravers \cite{1983AnPhy.150..455E,ETM1983}, following earlier work by \cite{ehlers:1971,ellis1971,thorne:1980,thorne:1981} (see also  \cite{Maartens:1994,stoeger:1995,Challinor:1998aa,Gebbie:1998fe,Maartens:1998xg,Maartens:1999mf,Lewis:2002nc,Tsagas:2007yx,Clarkson:2010uz,Clarkson:2012bg,Ellis:2012} for subsequent work).

At each event
$x^a$, the one-particle distribution function ${\sf f}(x^a,p^b)$ determines the number of particles seen by an observer in the rest-space volume element $\ud{\cal V}$ at $x^a$, whose geodesic 4-momenta $p^a$ lie in the mass-shell element $\ud {\cal P}$ about $p^b$:
\begin{align}\label{fn}
 \ud {\cal N}={\sf f}(x^a,p^b)(-u_c p^c)\,\ud {\cal V}\,\ud {\cal P}\,.
\end{align}
Here $u^c$ is the 4-velocity of the observer: $\ud {\cal N}$, ${\sf f}$ and $p^a$ are independent of the choice of $u^c$. A physical choice  is made below. The mass shell in momentum space is $p^a p_a=-m^2$, where $m$ is the particle mass. The relativistic 3-momentum is $\lam^a=\lam e^a$, with $e^a e_a=1$ and $e^a u_a=0$, $\ud\Omega=\ud^2e^a$ and $E=-u_c p^c$ is the particle energy, all as measured by the observer with 4-velocity $u^a$. The covariant momentum-space volume element is $\ud {\cal P}=\lam^2\ud\lam\ud\Omega/E=\lam\ud E\ud\Omega$. Note that $\lam\ud\lam=E\ud E$, which follows from $E^2-\lam^2=m^2$. In detail:
\begin{align}
\label{1}
 p^a&=Eu^a+\lam^a,\qquad E=-u_a p^a,\qquad \lam^a=\lam e^a,
 \notag\\
 \lam&=m\gamma(v)v,\qquad~ E=m\gamma(v),
 \qquad v^a=ve^a,\qquad \lam=Ev \,.
\end{align}
The Vlasov equation expresses conservation of the phase-space density of particles:
\begin{align}\label{vl}
 \mathcal{L}[{\sf f}]\equiv \Big(p^a\frac{\up }{\up x^a}-\Gamma^a_{bc}p^bp^c\frac{\up }{\up p^a}\Big)\, {\sf f}=0 \,.
\end{align}

Covariant relativistic kinetic theory expands ${\sf f}$ in tensor multipoles, which are a covariant equivalent of the coordinate-dependent $Y_{\ell m}$ \cite{1983AnPhy.150..455E,ETM1983}:
\begin{align}
 {\sf f}(x,E,e^a)=F(x,E)+F_a(x,E)e^a+F_{ab}(x,E)e^ae^b+\cdots
\end{align}
The multipoles are orthogonal to $u^a$ and are given by integrals over directions in momentum space:
\begin{align}
\label{3}
F_{a_1a_2\cdots a_\ell}=
\frac{(2\ell+1)!}{4\pi(\ell!)^2 2^\ell}\int\ud\Omega\,{\sf f}\,
e_{\langle a_1}e_{a_2}\cdots e_{a_\ell\rangle}\,,
\end{align}
where angled brackets $\langle a_1\cdots\rangle$ denote the projected (i.e. orthogonal to $u^a$), symmetric and tracefree part.
Note that the monopole is the average over directions $e^a$:
\begin{align}
    F= \frac{1}{4\pi} \int\ud\Omega\,{\sf f} \equiv \la {\sf f} \ra_{e}\,.
    \label{monf}
\end{align}
Integrals over the unit directions $e^a$  obey the identity
\begin{align}
\label{3x}
 \frac{\ell+1}{4\pi}\int\ud\Omega\,e^{a_1}e^{a_2}\cdots e^{a_\ell}
 =\begin{cases}
 0&\text{for $\ell$ odd},\\
 h^{(a_1a_2}h^{a_3a_4}\cdots h^{a_{\ell-1}a_\ell)}&\text{for $\ell$ even},
 \end{cases}
\end{align}
where round brackets $( a_1\cdots)$ denote the symmetric part.

\subsection{Energy-momentum tensor}
\label{s2.1}

The macroscopic energy-momentum tensor is an average over the microscopic distribution function, given by the integral
\begin{align}
\label{tab}
 T^{ab}=\int\ud {\cal P}\,p^ap^b\,{\sf f}
 =\rho u^au^b+Ph^{ab}+2q^{(a}u^{b)}+\pi^{ab}.
\end{align}
This implies integral expressions for the dynamical quantities. We give these integrals equivalently as energy and velocity integrals, using $\ud E=m\gamma^3v\ud v$:
\begin{align}
 \rho&=4\pi\int_m^\infty\ud E\,E^2\lam F
 ~~=4\pi m^4\int_0^1\ud v\,\gamma^6v^2F \,,
 \label{i1}\\
 P&=\frac{4\pi}{3}\int_m^\infty\ud E\,\lam^3F
~ ~~~=\frac{4\pi}{3}m^4\int_0^1\ud v\,\gamma^6v^4F \,,
 \label{i2}\\
 q_a&=\frac{4\pi}{3}\int_m^\infty\ud E\,E\lam^2F_a
 ~=\frac{4\pi}{3}m^4\int_0^1\ud v\,\gamma^6v^3F_a \,,
 \label{i3}\\
 \pi_{ab}&=\frac{8\pi}{15}\int_m^\infty\ud E\,\lam^3F_{ab}
 ~~\,=\frac{8\pi}{15}m^4\int_0^1\ud v\,\gamma^6v^4F_{ab}\,.
 \label{i4}
\end{align}
We see that $P$, $q_a$ and $\pi_{ab}$ are higher order in particle speed than $\rho$, and are therefore smaller for cold dark matter. We can write this indicatively as
\begin{align}\label{ovp}
 P,\pi_{ab}\sim O(v^2\rho),\qquad q_a\sim O(v\rho),
\end{align}
bearing in mind that the dynamical quantities are integrals over the full velocity distribution. This
approximation is made precise in \autoref{closure}.

The suppression of the higher moments is supplied by powers of particle speed; coldness alone does not require the angular coefficients $F_{a_1a_2\cdots a_\ell}$ themselves to decrease with $\ell$.
{The    coldness of dark matter} thus allows for a hierarchy truncation ordered by velocity weight. We truncate at the quadrupole to produce a self-consistent effective fluid model  in \autoref{s4}. The conditions controlling this truncation are discussed below and in \autoref{appa} and \autoref{appa2}.

The dust model is the monokinetic, distributional limit of the Vlasov description.
Dust is not a regular pointwise limit within the space of smooth phase-space distributions, and it ceases to provide a valid single-valued velocity field after stream crossing.

The kinematic quantities of the  matter-frame
4-velocity are given by its covariant derivative:
\begin{align}
 \nabla_bu_a=\frac13\Theta h_{ab}+\sigma_{ab}+\omega_{ab}-A_au_b \,,
\end{align}
where $h_{ab}=g_{ab}+u_au_b$ projects into the rest-space at each event, $\Theta=\nabla_au^a$ is the volume expansion rate of the $u^a$ flow-lines, $\si_{ab}=\si_{\langle ab\rangle}$ is their shearing rate, and $A_a=u^b\nabla_b u_a$ is their 4-acceleration, with $A_a u^a=0$. The vorticity is $\om_{ab}=h^c_b\, h^d_a\, \nabla_{[c} u_{d]} $, where square brackets denote the antisymmetric part. (Note that this definition keeps the order of indices on $u_{a;b}$\,, $\si_{ab}$ and $\om_{ab}$ the same, and consequently $\om_{ab}$ has the opposite sign of the Newtonian vorticity.)

\subsection{The matter frame}
\label{s2.2}

The energy-momentum tensor $T^{ab}$ is a physical tensor, independent of the choice of $u^a$, although its decomposition into $\rho$, $P$, $q_a$ and $\pi_{ab}$ changes with that choice. We make a physically motivated choice -- which we call the matter-frame 4-velocity -- by defining $u^a$ so that there is no relative matter energy flux:
\begin{align}
\label{14}
 \text{\sf matter frame:~~ no energy flux measured by}~ u^a ~ \text{\sf observers}
 \quad\Leftrightarrow\quad q_a=0.
\end{align}
This choice also simplifies the form of $T^{ab}$ and it then follows that $u^a$ is an eigenvector of $T^{ab}$.

The Vlasov equation \autoref{vl} enforces energy-momentum conservation, for any choice of $u^a$, since
\begin{align}
 \nabla_bT^{ab} = \int\ud {\cal P}\,p^a{\cal L}[{\sf f}]=0 \,.
\end{align}
This produces the exact, nonlinear energy and momentum conservation laws:
\begin{align}
 \dot\rho+(\rho+P)\Theta+\si^{ab}\pi_{ab}
 &=-2A^aq_a-\cd^aq_a \,,
 \label{ec}\\
 \cd_aP+(\rho+P)A_a+\cd^b\pi_{ab}+A^b\pi_{ab}
 &=-(\dot q_a)_{\pe}-\frac43\Theta q_a
 -\big(\sigma_{ab}+\omega_{ab}\big)q^b \,.
 \label{mc}
\end{align}
Here the overdot is the covariant time derivative along $u^a$ and $\pe$ denotes projection into the $u^a$ rest space. The covariant spatial derivative  $\cd_a$  satisfies \cite{Maartens:1998xg,Clarkson:2010uz}
\begin{align}
 \cd_aS^{b\cdots}_{\cdots c}
 &=h_a{}^dh_e{}^b\cdots h_c{}^f\nabla_dS^{e\cdots}_{\cdots f}\,,
 \qquad \cd_ah_{bc}=0\,.
\end{align}
In the matter frame, $q_a=0$ and the right-hand sides of the equations vanish. Consequently, momentum conservation turns from an evolution equation for $q_a$ into a constraint that shows how pressure gradients generate 4-acceleration in the average 4-velocity.

In the dust model of matter, the matter-frame 4-velocity is geodesic. A consequence of including velocity dispersion in the kinetic model of matter is that, in general, the matter-frame 4-velocity is no longer geodesic. There is an effective 4-acceleration $A_a=u^b\nabla_bu_a$ of the matter-frame 4-velocity, arising from averaging over velocity dispersion. In \autoref{s4} we derive the exact nonlinear expression for this 4-acceleration.

Individual galaxies are assumed to follow geodesics, so how does this relate to the non-geodesic average 4-velocity $u^a$?
\begin{itemize}
    \item
Firstly, cosmological survey observations impose an effective averaging over velocity dispersion, due to the huge distances involved and the resolution limits of telescopes.
\item
Secondly, since observations typically count galaxies,
number-weighted observations motivate comparing the particle-number
frame with the matter energy frame. These frames agree through second
velocity order for the same cold collisionless distribution.
This is shown as follows.
\end{itemize}

The  particle 4-current density is defined as
\begin{align}
 N^a=\int\ud {\cal P}\,p^a\,{\sf f}=nu^a+r^a,
\end{align}
where the particle number density and its flux relative to the matter frame are
\begin{align}
 n&=4\pi\int_m^\infty\ud E\,\lam EF
 \,=4\pi m^3\int_0^1\ud v\,\gamma^5v^2F,
 \label{ra}\\
 r_a&=\frac{4\pi}{3}\int_m^\infty\ud E\,\lam^2F_a
 =\frac{4\pi}{3}m^3\int_0^1\ud v\,\gamma^5v^3F_a.
\end{align}
The Vlasov equation again leads to a conservation law, in this case for the 4-current density:
\begin{align} \label{ndot}
 \nabla_aN^a=\int\ud {\cal P}\,{\cal L}[{\sf f}]=0
 \quad\Rightarrow\quad
 \dot n+\Theta n+\cd_ar^a+A_ar^a=0 \,.
\end{align}
The matter frame is defined by no energy-density flux,
$q^a=0$, while the frame with no number-density flux is defined by $r^a=0$. (The exact covariant relations between dynamical quantities in these frames, for a Boltzmann gas, are given in \cite{Maartens:1994}; see also \cite{Lewis:2002nc}.) The frames do not in general coincide, so that the matter frame will in general see a nonzero number-density flux, $r^a\ne0$. By \autoref{ra}, we find that, for a general 4-velocity,
\begin{align}
\label{mra}
mr_a &=\frac{4\pi}{3}m^4\int_0^1\ud v\,
 \gamma^5v^3F_a \,.
\end{align}
This means that $mr_a$ is $O(v\rho)$ -- the same as $q_a$. However, using \autoref{i3} and \autoref{mra}, we find that the difference between the two is smaller:
\begin{align}
\label{qmr}
 q_a-mr_a &=\frac{4\pi}{3}m^4\int_0^1\ud v\,
 (\gamma^6-\gamma^5)v^3F_a
 \notag\\
 &= \frac{2\pi}{3}m^4\int_0^1\ud v\,
 \gamma^5 v^3F_a \, \big[v^2 + O(v^4) \big] \,.
\end{align}
It follows that $q_a-mr_a$ is $O(v^3\rho)$ and may be neglected when we apply our approximation to the Vlasov hierarchy.  Therefore:
\begin{align}\label{frames}
\text{\sf In a cold collisionless gas, the energy and particle-number frames  agree up to}~ O(v^2\rho).
\end{align}
Identifying this common frame with the galaxy population additionally requires
negligible tracer velocity bias.

\section{Vlasov hierarchies}
\label{appa}

The distribution function  may be expanded in multipoles $F_{a_1a_2\cdots a_\ell}$, as in \autoref{3}. Then these multipoles obey an infinite hierarchy of coupled equations when the Vlasov equation is expanded. This hierarchy in phase space is given in \autoref{s3.1} below.

On the other hand, the macroscopic dynamics of the gas is governed by energy-integrated multipoles, as in \autoref{tab}. The infinite hierarchy of integrated multipoles in spacetime is considered in \autoref{appa2} below.

Both of these hierarchies were presented in the pioneering papers by Ellis, Matravers and Treciokas \cite{1983AnPhy.150..455E,ETM1983}.

\subsection{Hierarchy of the Vlasov equation in phase space}
\label{s3.1}

The general covariant multipole form of the collisionless Vlasov equation is given in Ellis, Matravers and Treciokas \cite{1983AnPhy.150..455E} (see \cite{Lewis:2002nc,Tsagas:2007yx,Ellis:2012} for more accessible versions).

First, we use the fact that the particle 4-momentum $p^a=Eu^a+\lam e^a$ is geodesic, so that $p^b\nabla_b p^a=0$, which implies
 that the 3-momentum and its direction evolve as
\begin{align}
 \frac{1}{E}p^b\nabla_b\lam &=-EA_ae^a-\lam\Big(\frac13\Theta+\si_{ab}e^ae^b\Big),
 \label{ddl}\\
 \frac{1}{E} \big( p^b\nabla_b e^a\big)_\pe &=-\frac{E}{\lam}\big(A^a-A_be^be^a\big)
 +\big(\si_{bc}e^be^c\big)e^a
 -\big(\si^{ab}+\om^{ab}\big)e_b \,.
 \label{dde}
\end{align}
\autoref{dde} is orthogonal to the propagation direction $e^a$, so that this equation describes gravitational lensing of individual particle flow-lines in the particle screen space, orthogonal to $u^a$ and $e^a$.

The Vlasov  \autoref{vl} can be rewritten as
\begin{align}
 \sum_{\ell=0}^\infty {\cal L}\Big[
 F_{{\cal A}_\ell}(x,E)e^{{\cal A}_\ell}\Big]=0\,, \qquad \text{with}\quad {\cal A}_\ell\equiv a_1a_2\cdots \,a_\ell \,,
\end{align}
where the multipoles obey $F_{{\cal A}_\ell}=F_{\langle{\cal A}_\ell\rangle}$.
Together with \autoref{ddl} and \autoref{dde}, this delivers a formula that generates the infinite Vlasov hierarchy \cite{1983AnPhy.150..455E}:
\begin{align}
 &E\big(\dot F_{{\cal A}_\ell}\big)_\pe
 -\frac{\lam^2}{3}\up_EF_{{\cal A}_\ell}\Theta
 +\frac{\ell+1}{2\ell+3}\lam\cd^aF_{a{\cal A}_\ell}
 +\lam\cd_{\langle a_\ell}F_{{\cal A}_{\ell-1}\rangle}
 \notag\\
 &+\ell EF^a{}_{\langle{\cal A}_{\ell-1}}
 \om_{a_\ell\rangle a}
 +\left[(\ell-1)\frac{E^2}{\lam}F_{\langle{\cal A}_{\ell-1}}
 -\lam E\up_EF_{\langle{\cal A}_{\ell-1}}\right]A_{a_\ell\rangle}
 \notag\\
 &-\frac{\ell+1}{2\ell+3}\bigg[(\ell+2)\frac{E^2}{\lam}
 F_{a{\cal A}_\ell}+\lam E\up_EF_{a{\cal A}_\ell}\bigg]A^a
 \notag\\
 &-\frac{\ell}{2\ell+3}\Big[3EF_{a\langle{\cal A}_{\ell-1}}
 +2\lam^2\up_EF_{a\langle{\cal A}_{\ell-1}}\Big]
 \si_{a_\ell\rangle}^{~~~a}
 \notag\\
 &-\frac{(\ell+1)(\ell+2)}{(2\ell+3)(2\ell+5)}
 \Big[(\ell+3)EF_{ab{\cal A}_\ell}
 +\lam^2\up_EF_{ab{\cal A}_\ell}\Big]\si^{ab}
 \notag\\
 &+\Big[(\ell-2)EF_{\langle{\cal A}_{\ell-2}}
 -\lam^2\up_EF_{\langle{\cal A}_{\ell-2}}\Big]
 \si_{a_{\ell-1}a_\ell\rangle}=0 \,.
 \label{aa}
\end{align}
Here $\up_E\equiv\up/\up E$. This form of the covariant relativistic Vlasov equation, which is nonperturbative and without assumptions on the spacetime, shows explicitly how the moments of the distribution function couple to the kinematic quantities of $u^a$ at all multipole orders.

The first three moments of the infinite Vlasov hierarchy are:
\begin{align}
\bm{\ell=0}:\quad &
 E\dot F-\frac13\lam^2\Theta\frac{\up F}{\up E}
 -\frac13\frac{E}{\lam}\frac{\up}{\up E}\Big(\lam^2A^aF_a\Big)
 +\frac13\lam\cd^aF_a
 \notag\\
&-\frac{2}{15}\frac{1}{\lam}\frac{\up}{\up E}
 \Big(\lam^3\si^{ab}F_{ab}\Big)=0 \,,
\label{l0}\\
\bm{\ell=1}:\quad &
 E(\dot F_a)_\pe-\frac13\lam^2\Theta\frac{\up F_a}{\up E}
 +\lam\cd_aF-\lam E\frac{\up F}{\up E}A_a
 -\frac25\lam^{1/2}\frac{\up}{\up E}
 \Big(\lam^{3/2}\si_{ab}F^b\Big)
 \notag\\
&+E\om_{ab}F^b+\frac25\lam\cd^bF_{ab}
 -\frac25\frac{E}{\lam^2}\frac{\up}{\up E}
 \Big(\lam^3A^bF_{ab}\Big)
 \notag\\
&-\frac{6}{35}\frac{1}{\lam^2}\frac{\up}{\up E}
 \Big(\lam^4\si^{bc}F_{abc}\Big)=0 \,,
\label{l1}\\
\bm{\ell=2}:\quad &
 E(\dot F_{ab})_\pe-\frac13\lam^2\Theta\frac{\up F_{ab}}{\up E}
 +\lam\cd_{\langle a}F_{b\rangle}
 -E\lam^2\frac{\up}{\up E}
 \Big(\lam^{-1}A_{\langle a}F_{b\rangle}\Big)
 \notag\\
&-\lam^2\frac{\up F}{\up E}\si_{ab}
 -\frac47\lam^{1/2}\frac{\up}{\up E}
 \Big(\lam^{3/2}F_{c\langle a}\si_{b\rangle}^{~c}\Big)
 +2EF_{c(a}\omega_{b)}^{~c}
 \notag\\
&+\frac37\lam\cd^cF_{abc}
 -\frac37\frac{E}{\lam^3}\frac{\up}{\up E}
 \Big(\lam^4A^cF_{abc}\Big)
 -\frac{4}{21}\frac{1}{\lam^3}\frac{\up}{\up E}
 \Big(\lam^5\si^{cd}F_{abcd}\Big)=0 \,.
\label{l2}
\end{align}
These covariant Vlasov multipole equations are fully nonperturbative and apply in any spacetime for any $u^a$. They contain the conservation equations, but also much more information.

\subsection{Energy-integrated hierarchy in spacetime}
\label{appa2}

\autoref{aa} determines the distribution function multipoles $F_{{\cal A}_\ell}(x,E)$ at each energy. For macroscopic dynamics we need only their energy-integrated moments.
In the case of massless particles a single energy weight suffices \cite{1983AnPhy.150..455E} (see also \cite{stoeger:1995,Challinor:1998aa,Gebbie:1998fe,Maartens:1998xg}).
For massive particles it does not: the mass shell $\lam^2=E^2-m^2$ means that each angular multipole carries an independent tower of velocity weights. This requires a two-index hierarchy: the exact Boltzmann form was derived by Ellis, Matravers and Treciokas \cite{1983AnPhy.150..455E,ETM1983} and further developed by Lewis and Challinor \cite{Lewis:2002nc}.

Here we recast the Vlasov hierarchy in the notation of the main text and define the covariant dynamical multipoles,
\begin{align}
 \Pi^{(n)}_{{\cal A}_\ell}(x) \equiv M_\ell\int_m^\infty\ud E\,
 \lam^{\ell+1}E^{2-\ell}\,v^{2n}F_{{\cal A}_\ell} (x,E)
 =\Pi^{(n)}_{\langle{\cal A}_\ell\rangle}(x)\,,
\qquad M_\ell&\equiv\frac{4\pi\,2^\ell(\ell!)^2}{(2\ell+1)!} \,,
 \label{Pidef}
\end{align}
where the integers $\ell,n \ge 0$ and $v=\lam/E$.
We write $\Pi_{{\cal A}_\ell}\equiv\Pi^{(0)}_{{\cal A}_\ell}$. (The corresponding moments in \cite{Lewis:2002nc} are $J^{(i)}_{{\cal A}_\ell}$, with velocity weight $\ell+2i$, using different signature and normalization conventions.)
All integrals are assumed to converge, which requires suitable asymptotic behaviour of $F_{{\cal A}_\ell}$ as $E\to\infty$ \cite{ehlers:1971}.

The normalisation $M_\ell$ is fixed by \autoref{3}, so that the $n=0$ tower is the set of dynamical quantities of \autoref{tab} and their higher-multipole extensions:
\begin{align}
 \Pi=\rho,\qquad \Pi_a=q_a,\qquad \Pi_{ab}=\pi_{ab},
 \qquad \Pi_{abc},\ \Pi_{abcd},\ \ldots,
 \qquad \Pi^{(1)}=3P.
 \label{Pident}
\end{align}
Thus the isotropic pressure is not the monopole of the $n=0$ tower but the monopole of the $n=1$ tower.

Equivalently, in terms of the particle speed, using $E=m\gamma$, $\lam=m\gamma v$ and $\ud E=m\gamma^3v\ud v$, we obtain
\begin{align}
 \Pi^{(n)}_{{\cal A}_\ell}(x)
 =M_\ell\, m^4\int_0^1\ud v\,\gamma^6v^{\ell+2n+2}
 F_{{\cal A}_\ell}(x,m\gamma)\,.
 \label{Pidefv}
\end{align}
The power of $\gamma$ is $(\ell+1)+(2-\ell)+3=6$  and the power of $m$ is $(\ell+1)+(2-\ell)+1=4$, both independent of $\ell$ and $n$. The weight $\lam^{\ell+1}E^{2-\ell}$ in \autoref{Pidef} is precisely the one for which every member of the family carries the same measure $m^4\gamma^6\ud v$, with the entire $(\ell,n)$--dependence residing in the single power $v^{\ell+2n+2}$.
For $(\ell,n)=(0,0),(1,0),(2,0)$ and $(0,1)$, \autoref{Pidefv} reproduces the velocity integrals in \autoref{i1}--\autoref{i4}.

We multiply \autoref{aa} by $M_\ell\lam^{\ell+1}E^{1-\ell}v^{2n}$ and integrate over $m\leq E<\infty$, using $\ud\lam/\ud E=E/\lam$ and integrating by parts. This gives the exact nonlinear dynamical hierarchy \cite{1983AnPhy.150..455E,Lewis:2002nc}:
\begin{align}
 &\big(\dot\Pi^{(n)}_{{\cal A}_\ell}\big)_\pe
 +\frac13\left[(\ell+3+2n)\Pi^{(n)}_{{\cal A}_\ell}
 +(1-\ell-2n)\Pi^{(n+1)}_{{\cal A}_\ell}\right]\Theta
 \notag\\
 &+\cd^a\Pi^{(n)}_{a{\cal A}_\ell}
 +\frac{\ell}{2\ell+1}\cd_{\langle a_\ell}
 \Pi^{(n+1)}_{{\cal A}_{\ell-1}\rangle}
 +\ell\Pi^{(n)a}_{~~~~\langle{\cal A}_{\ell-1}}
 \om_{a_\ell\rangle a}
 \notag\\
 &+\frac{\ell}{2\ell+1}\left[(2\ell+1+2n)
 \Pi^{(n)}_{\langle{\cal A}_{\ell-1}}
 +(2-\ell-2n)\Pi^{(n+1)}_{\langle{\cal A}_{\ell-1}}\right]
 A_{a_\ell\rangle}
 \notag\\
 &+\left[2n\Pi^{(n-1)}_{a{\cal A}_\ell}
 +(2-\ell-2n)\Pi^{(n)}_{a{\cal A}_\ell}\right]A^a
 \notag\\
 &+\frac{\ell}{2\ell+3}\left[(2\ell+3+4n)
 \Pi^{(n)}_{a\langle{\cal A}_{\ell-1}}
 +2(1-\ell-2n)\Pi^{(n+1)}_{a\langle{\cal A}_{\ell-1}}\right]
 \si_{a_\ell\rangle}^{~~~a}
 \notag\\
 &+\left[2n\Pi^{(n-1)}_{ab{\cal A}_\ell}
 +(1-\ell-2n)\Pi^{(n)}_{ab{\cal A}_\ell}\right]\si^{ab}
 \notag\\
 &+\frac{\ell(\ell-1)}{(2\ell+1)(2\ell-1)}
 \left[(2\ell+1+2n)\Pi^{(n+1)}_{\langle{\cal A}_{\ell-2}}
 +(1-\ell-2n)\Pi^{(n+2)}_{\langle{\cal A}_{\ell-2}}\right]
 \si_{a_{\ell-1}a_\ell\rangle}=0 \,.
 \label{fluidhier}
\end{align}
Note that $\Pi^{(n)}_{{\cal A}_\ell}=0$ for $\ell<0$ or $n<0$. This hierarchy is nonperturbative, holds for any $u^a$ in any spacetime, and involves no truncation.

Some structural features are worth noting.
\begin{itemize}
    \item
The divergence term
\[ \cd^a\Pi^{(n)}_{a{\cal A}_\ell}\] has unit coefficient at every order, since  $(\ell+1)/(2\ell+3)$ in \autoref{aa} is exactly cancelled by $M_\ell/M_{\ell+1}$.
    \item
The coefficient $(1-\ell-2n)$ of the $\si^{ab}\Pi^{(n)}_{ab{\cal A}_\ell}$ coupling vanishes at $\ell=1$, $n=0$: this is the covariant reason that the octupole cannot contribute to momentum conservation, as required by \autoref{tab}. This is an exact result and is independent of the choice of $u^a$.
\item
Similarly, $(2-\ell-2n)$ vanishes at $\ell=2$, $n=0$, so no $A^c\Pi_{abc}$ term appears in the anisotropic-pressure equation.
    \item
The 4-acceleration and shear couple the $n$ towers downwards through the $2n\Pi^{(n-1)}$ terms, so the velocity-weight index is not a one-way hierarchy.
\end{itemize}

The first five equations in the hierarchy are as follows.
Setting $(\ell,n)=(0,0)$ and $(0,1)$ in \autoref{fluidhier} gives the energy and pressure equations; $(\ell,n)=(1,0)$ gives the momentum equation; the anisotropic-pressure equation is $(\ell,n)=(2,0)$ , and $(\ell,n)=(3,0)$ gives the octupole equation. Using the notation
\begin{align}
 {\cal P}\equiv\tfrac13\Pi^{(2)}\,,\quad~~
 {\cal Q}_a\equiv\Pi^{(1)}_a\,,\quad~~
 {\cal R}_{ab}\equiv\Pi^{(1)}_{ab}\,,\quad~~
 {\cal Q}^{(2)}_a\equiv\Pi^{(2)}_a\,,\quad~~
 {\cal S}^{(2)}_{abc}\equiv\Pi^{(1)}_{abc}\,,
 \label{auxdef}
\end{align}
we can find the first five exact energy-integrated evolution equations from \autoref{fluidhier}, which allow us to see explicitly how the second velocity order truncation emerges. We obtain:
\begin{align}
 \bm{(\ell,n)=(0,0)}:\quad
 &\dot\rho+(\rho+P)\Theta+\si^{ab}\pi_{ab} +\cd^aq_a+2A^aq_a
 =0\,,
 \label{fh0}\\[3mm]
 \bm{(\ell,n)=(0,1)}:\quad
 &\dot P+\frac53\Theta P+\frac23\si^{ab}\pi_{ab}
 +\frac23A^aq_a
 \notag\\
 &\qquad +\frac13\cd^a{\cal Q}_a
 -\frac13\Theta{\cal P}-\frac13\si^{ab}{\cal R}_{ab}=0\,,
 \label{fh0P}\\[3mm]
 \bm{(\ell,n)=(1,0)}:\quad
 &(\dot q_a)_\pe+\frac43\Theta q_a +\big(\si_{ab}+\om_{ab}\big)q^b
 \notag \\
 &\qquad +(\rho+P)A_a
 +\cd_aP+\cd^b\pi_{ab}
 +A^b\pi_{ab}
 =0\,,
 \label{fh1}\\[3mm]
 \bm{(\ell,n)=(2,0)}:\quad
 &(\dot\pi_{ab})_\pe+\frac53\Theta\pi_{ab}
 +2\pi_{c\langle a}\big(\si_{b\rangle}^{~c}+\om_{b\rangle}^{~c}\big)
 +2P\si_{ab} +2A_{\langle a}q_{b\rangle}
 \label{fh2}  \\ \notag
 &\qquad+\cd^c\Pi_{abc}+\frac25\cd_{\langle a}{\cal Q}_{b\rangle}
 -\si^{cd}\Pi_{abcd}-\frac13\Theta{\cal R}_{ab}
 -\frac47{\cal R}_{c\langle a}\si_{b\rangle}^{~c}
 -\frac25{\cal P}\si_{ab}=0\,,
  \\[3mm]
 \bm{(\ell,n)=(3,0)}:\quad
 &(\dot\Pi_{abc})_\pe+2\Theta\Pi_{abc}
 +3\Pi_{d\langle ab}\big(\si_{c\rangle}^{~d}+\om_{c\rangle}^{~d}\big) +3A_{\langle a}\pi_{bc\rangle}
 \notag\\
 &\qquad+\frac65{\cal Q}_{\langle a}\si_{bc\rangle}
 +\cd^d\Pi_{abcd}+\frac37\cd_{\langle a}{\cal R}_{bc\rangle}
 -A^d\Pi_{abcd}
 -2\si^{de}\Pi_{abcde}
 \notag\\
 &\qquad -\frac23\Theta{\cal S}^{(2)}_{abc}
 -\frac43{\cal S}^{(2)}_{d\langle ab}\si_{c\rangle}^{~d}
 -\frac37A_{\langle a}{\cal R}_{bc\rangle}
 -\frac{12}{35}{\cal Q}^{(2)}_{\langle a}\si_{bc\rangle}=0\,.
 \label{fh3}
\end{align}
Note that these equations apply for any choice of 4-velocity.
\autoref{fh0} and \autoref{fh1} are exactly the conservation laws \autoref{ec} and \autoref{mc}, now obtained as the two lowest members of a single formula. In the matter frame $q_a=0$, and \autoref{fh1} becomes the acceleration constraint
(see \autoref{s4}).
\autoref{fh1} and \autoref{fh2} generalise the results of \cite{Maartens:1999mf}, which neglects products of multipoles. In \autoref{fh3}, when we apply the quadrupole truncation, a term
$3A_{\langle a} \pi_{bc\rangle}$ appears to survive. However, this term is fourth velocity order and therefore vanishes. Therefore the octupole equation imposes no additional condition
at second velocity order.

\subsection{Velocity ordering and the second-moment closure}
\label{closure}

Because the measure $m^4\gamma^6\ud v$ in \autoref{Pidefv} is common to every member of the family, the ratio of any two integrands is a pure power of $v$ times a ratio of distribution multipoles, and the order-counting
\begin{align}
 \Pi^{(n)}_{{\cal A}_\ell}=O\big(v^{\ell+2n}\rho\big) \,,
 \label{Piorder}
\end{align}
is exact in that sense: no $\gamma$ factors survive in the relative ordering, and the $\gamma^6$ that is common to all members contributes only the overall $1+3v^2+O(v^4)$ correction already exhibited in \autoref{i1}. Thus $\ell+2n$ is the velocity order of each integrated multipole. This organises the hierarchy:
\begin{align}
 \text{\sf
at}~~ O(v^2\rho)~~ \text{\sf the complete set is}~ (\ell,n)=(0,0),(1,0),(0,1),(2,0) \Leftrightarrow \{\rho,q_a,P,\pi_{ab}\} \,,
\end{align}
and no others.
These carry $1+3+1+5=10$ independent components, precisely the content of $T^{ab}$: the $\ell+2n\leq2$ sector of \autoref{Pidef} is \autoref{tab}, with $\ell$ indicating angular structure and $n$ indicating the trace parts.

The next order $O(v^3\rho)$ introduces exactly two new objects, the octupole $\Pi_{abc}$ and the vector ${\cal Q}_a$, and $O(v^4\rho)$ introduces $\Pi_{abcd}$, ${\cal R}_{ab}$ and ${\cal P}$. The truncation used in \autoref{s2} (see \autoref{ovp}) is therefore the statement that
\begin{align}\label{trunc}
\text{\sf every}~~\Pi^{(n)}_{{\cal A}_\ell} ~~ \text{\sf with}~~\ell+2n\geq3~~\text{\sf is set to zero.}
\end{align}
Thus \autoref{fh0} and \autoref{fh1} are kept,
while in \autoref{fh0P} and \autoref{fh2} we discard the second line.  \autoref{fh3} reduces to $0=0$.
Note that this is a {\em single} condition on $\ell+2n$, not separately on $\ell$: discarding the angular octupole $\Pi_{abc}$ while retaining ${\cal Q}_a$, or conversely, is not a consistent velocity-order truncation.
This is the $n_*=2$ velocity-weight truncation used  in \cite{Lewis:2002nc}. (Its controlled large-scale condition on an FLRW background is $kv_{\rm rms}\ll aH$,
as shown in
the companion perturbative paper \cite{Clarkson:2026xxx}.)

The natural dimensionless velocity moments are  obtained by normalising with the energy density, following Maartens, Gebbie and Ellis \cite{Maartens:1998xg}:
\begin{align}
 \tau^{(n)}_{{\cal A}_\ell}\equiv
 \frac{\Pi^{(n)}_{{\cal A}_\ell}}{\rho}
 =\llangle v^{2n}v_{\langle a_1}v_{a_2}\cdots
 v_{a_\ell\rangle}\rrangle \qquad \text{where} \qquad
  \llangle {\cal X}\rrangle\equiv\frac1\rho\int\ud E\,\ud\Omega\,
 \lam E^2{\cal X}\,{\sf f}\,.
 \label{tauobs}
\end{align}
Here $v^a=ve^a$ and $\llangle \cdots\rrangle$ is the energy-weighted average over the local momentum distribution. \autoref{tauobs} is exact. It reduces to the number-weighted average over velocities for $v\ll1$.
In particular,
\begin{align}
 \tau^{(0)}_a&=\llangle v_a\rrangle=\frac{q_a}{\rho},
 &\tau^{(1)}&=\llangle v^2\rrangle=\frac{3P}{\rho},
 &\tau^{(0)}_{ab}&=\llangle v_{\langle a}v_{b\rangle}\rrangle
 =\frac{\pi_{ab}}{\rho},
 \notag\\
 \tau^{(1)}_a&=\llangle v^2v_a\rrangle=\frac{{\cal Q}_a}{\rho},
 &\tau^{(0)}_{abc}&=\llangle v_{\langle a}v_bv_{c\rangle}\rrangle
 =\frac{\Pi_{abc}}{\rho}.
 \label{tauex}
\end{align}

\vspace*{0.5cm}
\section{Self-consistent effective fluid}
\label{s4}

Using \autoref{fh0}--\autoref{fh3}, with $q_a=0$, and imposing \autoref{trunc}, we have a  truncated Vlasov hierarchy in the matter frame, where all higher-order terms and equations are self-consistently neglected:
\begin{align}
 \dot\rho+(\rho+P)\Theta+\si^{ab}\pi_{ab}&=0\,,
 \label{dr}\\
 \dot P+\frac53\Theta P+\frac23\si^{ab}\pi_{ab}&=0\,,
 \label{dp}\\
 (\dot\pi_{ab})_\pe+\frac53\Theta\pi_{ab}
 +2\pi^c_{~\langle a}\big(\si_{b\rangle c}+\om_{b\rangle c}\big)
 +2P\si_{ab}&=0\,,
 \label{dpi}\\
(\rho+P)A_a+\cd_aP+\cd^b\pi_{ab}+A^b\pi_{ab}&=0\,.
 \label{da}
\end{align}
We reiterate that no background spacetime is assumed. Our approximation is fully nonperturbative in spacetime inhomogeneity and in the kinematic and dynamical quantities, but is truncated in velocity order. (Recall that \autoref{dr} and \autoref{da} are exact.)

This delivers a closed system at the cost of removing the sourcing of $\pi_{ab}$ by the octupole $\Pi_{abc}$ and the chain of sourcing by higher-order multipoles, given in \autoref{fluidhier}.
Truncation clearly does not fully capture the physics of multi-streaming. It retains the leading anisotropic velocity-dispersion corrections to dust on scales where the velocity hierarchy is controlled. Truncating at higher order would improve the description on nonlinear scales, where higher cumulants can be important \cite{Lewis:2002nc,Garny:2025ovs}.

Note that the closed system \autoref{dr}--\autoref{da} does not generate post-crossing dispersion from a cold initial state $P_{\rm in} = 0= \pi^{\rm in}_{ab} $, since it is truncated after $O(v^2\rho)$. Instead, it describes an already velocity-dispersed distribution in a controlled approximation.

There is also an inherent limit on the validity of the collisionless model, even with an infinite hierarchy: it cannot include thermal baryonic effects, which can strongly affect the evolution of matter clustering on small enough scales.

A self-consistent effective fluid description is based on \autoref{dr}--\autoref{da}. However, we should be careful to understand this as an analogy: a collisionless gas is very different from a hydrodynamic fluid, which is collision dominated. In particular, the role and nature of pressure is different. The collisionless pressure is not thermal, like hydrodynamic pressure, but is the leading isotropic stress induced by velocity dispersion. The interaction between particles is {\em purely gravitational in the Vlasov model}, whereas particles in a hydrodynamic fluid interact strongly via collisions, as well as via gravity.

\subsection{Kinetic pressure}
\label{s4.1}

In a dust model, $\rho_{\rm dust}=mn$. The isotropic  pressure generated by velocity dispersion modifies this relation. Using  \autoref{i1}, \autoref{i2},  \autoref{ra} and \autoref{auxdef}, we find that
\begin{align} \label{mn}
    \rho - mn &= \frac32 P +  \frac12 \pi m^4 \int_0^1 \ud v\,\gamma^6\, v^2\, F\, \big[v^4 +O(v^6)\big]
    \notag\\
    &= \frac32 P +  \frac18 \Pi^{(2)} + O(v^6\rho)\,,
\end{align}
in agreement with \cite{Maartens:1999mf}. The second term on the right is $O(v^4\rho)$ and is neglected in our approximation, \autoref{trunc}.

\autoref{dr} and \autoref{dp} imply that the isotropic pressure  and the density satisfy the evolution equation
\begin{equation}\label{pr}
 \Big(\rho-\frac32P\Big)^{\!\displaystyle{\bm\cdot}}
 +\Theta \Big(\rho-\frac32P\Big)=0\,,
\end{equation}
where the $\si^{ab}\pi_{ab}$ terms cancel out.
By \autoref{mn}, $mn= \rho-3P/2$ at $O(v^2\rho)$.
Since  $r_a=0$ in the matter frame at this order, by \autoref{qmr}, it follows that \autoref{pr}
is equivalent to number density conservation, $\dot n + \Theta n=0$, at $O(v^2\rho)$.
We note that the exact form of \autoref{pr} is
\begin{align} \label{pre}
  \Big(\rho-\frac32P\Big)^{\!\displaystyle{\bm\cdot}}
 +\Theta\Big(\rho-\frac32P\Big) =
 \frac12 \cd^a {\cal Q}_a - \frac12 \Theta {\cal P} - \frac12 \si^{ab} {\cal R}_{ab\,.}
\end{align}

Both $\rho$ and $P$ evolve under the influence of the anisotropic pressure generated by velocity dispersion. The isotropic pressure $P$ and anisotropic pressure $\pi_{ab}$ source each other via the shear anisotropy $\si_{ab}$. Without the truncation \autoref{trunc} of the Vlasov hierarchy, these variables would no longer form a closed subsystem, since higher moments would provide additional sources.

The extension of standard perturbation theory developed in the effective field theory of large-scale structure introduces nonperturbative counterterms after averaging over nonlinear scales. These terms can be understood in terms of an effective imperfect fluid \cite{Baumann:2010tm,Ivanov:2026nlf}, with effective bulk and shear viscosities which need to be calibrated by simulations.
By contrast, our evolution equations \autoref{dp} and \autoref{dpi} are determined self-consistently by  the Vlasov equation, within the controlled closure \autoref{trunc}.

\subsection{4-acceleration}
\label{s4.2}

The exact \autoref{da} can be rewritten as
\begin{align}\label{da2}
 \big[(\rho+P)h_{ab}+\pi_{ab}\big]A^b
 =-\cd^b\big(Ph_{ab}+\pi_{ab}\big).
\end{align}
Since $(\rho+P)h_{ab}+\pi_{ab}$ is invertible, we can solve this relation  for $A^a$ using the Cayley--Hamilton theorem:
\begin{align}\label{dax}
\notag \text{\sf exact nonlinear} & \text{\sf ~4-acceleration  (matter frame):}
\\
 A_a=(\rho+P)^{-1} & \left[1-\frac12\tilde\pi_{de}\tilde\pi^{de}
+\frac13\tilde\pi_{de}\tilde\pi^e{}_{f}\tilde\pi^{fd}\right]^{-1}
 \notag\\
 &\times\left[\tilde\pi_{ab}-\tilde\pi_{ac}\tilde\pi^c{}_{b}
 -\left(1-\frac12\tilde\pi_{de}\tilde\pi^{de}\right)h_{ab}\right]
 \cd_g\big(Ph^{bg}+\pi^{bg}\big),
\end{align}
where we defined the dimensionless quantity $\tilde\pi_{ab}\equiv\pi_{ab}/(\rho+P)$. This shows explicitly how the pressure gradients and anisotropic pressure, which are generated by averaging over velocity dispersion, produce a deviation from geodesic behaviour of the matter frame.

The exact results \autoref{da2} and \autoref{dax} reduce at $O(v^2\rho)$ to
\begin{align}\label{da3}
 A_a=-(\rho+P)^{-1}\big(\cd_aP+\cd^b\pi_{ab}\big).
\end{align}
Note that the 4-acceleration arises purely from velocity dispersion, unlike the kinematic quantities $\Theta$ and $\si_{ab}$. In \autoref{s4.3} we show that $\om_{ab}$ can also be generated by velocity dispersion.

\subsection{Vorticity generation}
\label{s4.3}

In the case of dust there is no source for vorticity (see e.g. \cite{Ellis:1971pg,Ellis:2012,Jelic-Cizmek:2018gdp,Umeh:2023lbc}). The question then arises: is vorticity generated by multi-streaming in a collisionless gas? The covariant vorticity evolution equation \cite{Maartens:1998xg,Ellis:2012},
\begin{align}
 (\dot\om_a)_\pe+\frac23\Theta\,\om_a-\si_{ab}\,\om^b
 =-\frac12\,{\rm curl}\,A_a,
 \label{vor}
\end{align}
is nonlinear and applies in any spacetime. The covariant spatial curl and the vorticity vector are defined by
\begin{align}
 {\rm curl}\,A_a\equiv\eta_{abc}\,\cd^bA^c,
 \qquad
 \om_a\equiv\frac12\eta_{abc}\,\om^{bc}~~
 \text{and}~~ \om_{ab}\equiv\eta_{abc}\,\om^{c},
\end{align}
where the vorticity vector is dual to the tensor, i.e., $\om_{ab}\,\om^b=0$, and thus is aligned along the local rotation axis. Here $\eta_{abc}=\eta_{abcd}u^d=\eta_{[abc]}$ is the spatial projection of the spacetime alternating tensor, $\eta_{abcd}=\eta_{[abcd]}$. \autoref{vor} shows that the 4-acceleration, which is generated by collective motion of the collisionless gas, is a source for vorticity. Vorticity generation in the full nonlinear case is governed by the exact relations \autoref{dax} and \autoref{vor}. In our truncated model, we use \autoref{da3} for the 4-acceleration.

Note that not all of the contributions to $A_a$ in \autoref{da3} can generate vorticity. Firstly, the curl of any spatial gradient, and in particular that of $P$, is nonzero in the presence of vorticity \cite{Maartens:1998xg,Ellis:2012}:
\begin{align}
 {\rm curl}\,\cd_aP=-2\dot P\om_a\,,
\end{align}
and similarly for $\cd_a \rho$. This reflects non-commutativity of spatial covariant
derivatives,
\begin{align*}
 \cd_{[a}\cd_{b]}P=-\dot P\om_{ab}\,,
\end{align*}
in the presence of vorticity -- since vorticity implies the absence of integrable
3-surfaces orthogonal to $u^a$. Spatial gradients on their own do not generate
vorticity; their curl reflects the presence of vorticity and affects its
evolution. However, if the two scalars have non-aligned gradients, i.e.
$\eta_{abc}\cd^b\rho\,\cd^cP\ne0$, then vorticity is generated.

\autoref{da3} shows that
\begin{align}\label{da4}
 {\rm curl}\,A_a={}&\frac{2\dot P}{\rho+P}\,\om_a
 -\frac{1}{\rho+P}\,{\rm curl}\,\cd^b\pi_{ab}
 \notag\\
 &+\frac{1}{(\rho+P)^2}\,\eta_{abc}\cd^b\rho \,
 \big(\cd^cP+\cd_d\pi^{cd}\big).
\end{align}
In general, ${\rm curl}\,\cd^b\pi_{ab}$ is nonzero and may also generate vorticity. The vorticity evolution \autoref{vor} becomes:
\begin{align}
\label{vor2}
 \notag &\text{\sf nonlinear vorticity evolution at}~ O(v^2\rho)~   \text{\sf(matter frame):}
\\
 &(\dot\om_a)_\pe+\Big(\frac23\Theta+\frac{\dot P}{\rho+P}\Big)\,\om_a
 -\si_{ab}\,\om^b
 \notag\\
 &\qquad=\frac{1}{2(\rho+P)}\,{\rm curl}\,\cd^b\pi_{ab}
 -\frac{1}{2(\rho+P)^2}\,\eta_{abc}\cd^b\rho \,
 \big(\cd^cP+\cd_d\pi^{cd}\big).
\end{align}
In summary, the  sources of vorticity on the right, at $O(v^2\rho)$, are the curl of the divergence of $\pi_{ab}$, the antisymmetric product of the density and pressure gradients, and the antisymmetric product of the density gradient and the divergence of $\pi_{ab}$.
At linear scalar order, these sources vanish -- although ${\rm curl}\,\cd^b\pi_{ab}$ has a linear vector mode.

\section{{Covariant cosmography}}
\label{s5}
{The covariant definition of the Hubble rate in any spacetime, together with its transformation to the boosted frame of the observer, was presented in \cite{Maartens:2023tib}, for the case of a dust model of matter. Here we generalise this to collisionless matter by including 4-acceleration. The Hubble rate, which depends on direction $n^a$, is defined covariantly in the matter frame $u^a$ by projecting the physical expansion tensor onto the past lightcone  of the observer \cite{Kristian:1966,MacCallum:1970,Maartens:2023tib}:
\begin{align} \label{hub}
    \mathbb{H}(n^a)& \circeq K^a K^b\,\nabla_a u_b
    \qquad\text{where}\qquad K^a= - u^a +n^a
    \notag\\
    & \circeq \frac13 \Theta -A_a\,n^a+\si_{ab}\,n^a n^b
    \notag\\
    &\circeq \langle \mathbb{H} \rangle  +\mathbb{H}_a\,n^a +\mathbb{H}_{ab}\, n^a n^b\,.
\end{align}
Here $\circeq$ denotes equality at the observer spacetime event o (`here and now').
{The monopole $\langle \mathbb{H} \rangle\equiv\langle \mathbb{H} \rangle_{n}$ is the observer's sky average.}
We use the dimensionless past-directed null vector $K^a=k^a/(u_b k^b)$, where $k^a$ is the past-directed photon 4-momentum -- in order to remove spurious frequency dependence in the Hubble rate.

Thus the matter-frame 4-acceleration is the intrinsic Hubble dipole, as measured in the matter frame. It vanishes for dust, but is generated by the isotropic and anisotropic pressure of collisionless matter through \autoref{da3}.
The covariant luminosity distance--redshift relation \cite{Kristian:1966,MacCallum:1970,Clarkson:1999zq,Hasse:1999,Umeh:2013,Clarkson:2010uz,Heinesen:2020bej,Maartens:2023tib} is valid  in any spacetime and given to $O(z^2)$ by
\begin{align}
 d_L(z,n^a)=\frac{z}{\mathbb H_{\rm o}(n^a)}
 \left[1+\frac12\big\{1-\mathbb Q_{\rm o}(n^a)\big\}z+O(z^2)\right].
 \label{eq:main-dLcos}
\end{align}
Here   ${\mathbb H}$ is given by \autoref{hub} and ${\mathbb Q}$ is the covariant deceleration parameter (defined below).
The matter-frame dipole is the 4-acceleration, providing a direct way of testing whether the breakdown of the dust cosmography can be explained by velocity dispersion.

Following \cite{Kalbouneh:2025jnp}, we probe anisotropy in the redshift-distance relation in the matter frame via
\begin{align}
    \eta(z,n^a)= \log_{10}\frac{z}{d_L(z,n^a)}
    -\Big\langle \log_{10}\frac{z}{d_L(z,n^a)}\Big\rangle\,,
    \label{eq:main-eta-def}
\end{align}
where  the angular monopole (the sky average) is removed.
In practice the matter frame is not known a priori and must be determined by observational data.
The observer 4-velocity  is
\begin{align}
   \tilde u^a \circeq \gamma(\w)\big( u^a+\w^a \big) \circeq u^a+\w^a +O(\w^2) \quad \text{with} \quad u^a \w_a\circeq 0\,,
\end{align}
where  we neglect $O(\w^2)$ terms since $\w\ll1$.
The velocity $\w^a$ of the observer relative to the matter frame is defined only at the observation event o. It can be determined along with the Hubble multipoles, following the method of \cite{Maartens:2023tib,Kalbouneh:2024szq,Kalbouneh:2024yjj,Kalbouneh:2025jnp}. In order to do this, we need in particular the Hubble parameter and its multipoles that are measured by the  boosted observer. We use the fact that $\nabla_a u_b$ is a physically defined tensor that is independent of the observer at event o. Therefore
\begin{align} \label{bhub}
    \tilde{\mathbb{H}}(\tilde n^a)&\circeq \tilde K^a \tilde K^b\,\nabla_a u_b
    \\ \notag
    & \circeq \Big[\frac13 \Theta +{\frac43} A_a \w^a\Big]
    -\Big[A_a +\frac23 \Theta\, \w_a + 2\si_{ab}\w^b \Big]\,\tilde n^a+\big[\si_{ab}+ A_{\langle a} \w_{b\rangle}\big] \,\tilde n^a \tilde n^b
    +O(\w^2)\,,
\end{align}
where  $\tilde K^a= - \tilde u^a +\tilde n^a$.
The reconstruction of cosmographic parameters using Cosmicflows-4 and upcoming data, following the method of \cite{Kalbouneh:2025jnp}, will allow the 4-acceleration $A_a$ to be estimated without assuming a background spacetime.  The companion paper \cite{Clarkson:2026xxx} specialises the kinetic system to perturbed FLRW spacetime and gives a theoretical estimate using Vlasov perturbation theory.

\subsection{Covariant deceleration parameter}
\label{5.1a}

The next order in \autoref{eq:main-dLcos} requires the covariant deceleration parameter
$\mathbb Q(n^a)$ \cite{MacCallum:1970,Maartens:2023tib,Heinesen:2020bej,Kalbouneh:2024yjj}.
We start by taking one more
derivative along the lightray:
\begin{align}
 \mathbb X(n^a)\equiv K^aK^bK^c\,\nabla_a\big(\nabla_{b}u_c\big)
 =\langle\mathbb X\rangle +\mathbb X_a\,n^a+\mathbb X_{ab}\,n^{\langle a} n^{b\rangle}
 +\mathbb X_{abc}\,n^{\langle a} n^b n^{c\rangle}\,.
 \label{X}
\end{align}
In the FLRW limit this is
$\langle\mathbb X\rangle=2H^2-\dot H=(q+3)H^2$, so that the covariant deceleration parameter can be defined by
\begin{align}
    \mathbb X=\big(\mathbb Q+3\big)\mathbb H^2 \,.
\end{align}
This $\mathbb Q$ reduces to the usual
deceleration parameter, as it must for consistency with \autoref{eq:main-dLcos}. The
decomposition in \autoref{X} terminates at $\ell=3$, and the reason generalises:
$\mathbb X$ carries three factors of $K^a$, each of which contributes at most one power of
$n^a$, so the $m$th member of the tower is a polynomial of degree
$m+2$ in $n^a$. Thus there is no
hexadecapole in
$\mathbb X$.

The deceleration parameter $\mathbb Q=
\mathbb X/\mathbb H^2-3$ can contain  higher multipoles in the general nonlinear case, because $\mathbb H^{-2}$ is not a finite
polynomial in $n^a$.
However these higher multipoles are induced by the lower multipoles in $\mathbb H$ -- and do not contain more independent information than $\mathbb H$ and $\mathbb X$. Consequently,  there is no independent hexadecapole in $d_L$ at $O(z^2)$.
This settles a discrepancy in the literature at the next order: \cite{Heinesen:2020bej}
finds $9+16+36$ independent components up to the jerk term,
whereas \cite{Kalbouneh:2024yjj}
find $9+16+25$, having discovered by direct computation that the $\ell=5$ multipole of the
jerk is not independent. In the variables of \autoref{X} the apparent high-$\ell$ multipoles
are artefacts of dividing by powers of $\mathbb H$, and the independent content at order
$z^m$ is $(m+3)^2$.

Evaluating \autoref{X} in a general spacetime, with no restriction on the kinematic or
dynamical quantities, we find that
\begin{align}
 \langle\mathbb X\rangle &=-\frac13\dot\Theta+\frac29\Theta^2
 +\frac23\si_{ab}\si^{ab}-\frac13\cd^aA_a+\frac43 A_a A^a\,,
 \\
 \mathbb X_a&=\frac13\cd_a\Theta+\frac25\cd^b\si_{ab}
 +(\dot A_a)_\pe-\frac43\Theta A_a
 -\frac{17}{5}\si_{ab}A^b{}+{}\om_{ab}A^b\,,
 \\
 \mathbb X_{ab}&=-(\dot\si_{ab})_\pe+\frac43\Theta\si_{ab}
 +2\si_{c\langle a}\si_{b\rangle}{}^c
 -\cd_{\langle a}A_{b\rangle}+A_{\langle a}A_{b\rangle}
 {}+2\si_{\langle a}{}^c\om_{cb\rangle}\,,
 \\
 \mathbb X_{abc}&=\cd_{\langle a}\si_{bc\rangle}-A_{\langle a}\si_{bc\rangle}\,.
 \label{Xmult1}
\end{align}
Setting $A_a=0=\om_{ab}$ reduces these expressions to the dust results of \cite{Maartens:2023tib}.
Converting to the deceleration variable of \cite{Heinesen:2020bej} reproduces the
multipoles given there (except for the coefficient of $\om_{ab}A^b$). The key point about our results, compared to previous work, is that we are able to determine the 4-acceleration self-consistently via the Vlasov model, as shown in \autoref{s4.2}.

{The Ricci identity provides exact covariant propagation and constraint equations for the shear
\cite{Ellis:1971pg,Maartens:1998xg}. In the matter frame, these are
\begin{align}
   (\dot\si_{ab})_\pe+\frac23 \Theta \si_{ab} &= -E_{ab} +4\pi G\, \pi_{ab} + \cd_{\la a} A_{b\ra} -\si_{c\la a} \si_{b\ra}^{~c}
   +A_{\la a} A_{b\ra} -\om_{\la a} \om_{b\ra}\,,
   \label{emcf}
   \\
   \cd^b\si_{ab}-\frac23\cd_a\Theta &= {\rm curl\,} \omega_a + 2\,\om_{ab}\,A^b -8\pi G\, \hat q_a \,.
   \label{emce}
\end{align}
Here $E_{ab}=C_{acbd}\,u^c u^d = E_{\la ab\ra}$ is the
electric Weyl tensor and $\hat q^a$ is the energy flux from radiation and other sources of the Einstein equations -- which we neglect for late-time local cosmography. }

\subsection{Linearised covariant analysis}
\label{s5.1}

We now linearise about FLRW, extending Maartens, Triginer and Matravers \cite{Maartens:1999mf}. In the covariant approach this means discarding
products of quantities that vanish in the background: $A_a$, $\si_{ab}$, $\om_{ab}$, $\pi_{ab}$ and $E_{ab}$ are first order because
their background values vanish, while $\cd_a\Theta$, $\cd_aP$ and $\cd_a\rho$  are first
order because the background is homogeneous \cite{Ellis:1998ct}.

{FLRW admits a collisionless distribution that is isotropic, ${\sf f}(x^a,E,e^b)=F(t,E)$.
In the background, $\bar\Theta=3H$ and isotropic kinetic stress generates a pressure $\bar P$.  \autoref{dr}, \autoref{dp} and \autoref{mn} give the background relations
\begin{align}\label{barr}
    \bar\rho =m \bar n +\frac32 \bar P
    = m \bar n_0\, a^{-3} +\frac32 \bar P_0 \,a^{-5}\,,
    \quad
    \frac{\bar P}{m\bar n}  = \frac{\bar P_0}{m\bar n_0}\, a^{-2}\,.
\end{align}
We  neglect radiation and assume that dark energy is a cosmological constant, which only affects the background -- as is appropriate for late-time local cosmology.
The Ricci identity produces the nonlinear Raychaudhuri evolution equation
\begin{align}
    \dot\Theta +\frac13 \Theta^2 + 4\pi G (\rho+3P) -\Lambda -\cd^aA_a = -\si_{ab} \si^{ab}+ 2\om_{a} \om^{a} + A_a A^a \,.
\end{align}
Then the Einstein field equations produce the $\delta\rho + 3\delta P$ term in
the linearised Raychaudhuri equation:
\begin{align}
    (\delta\Theta )^{\dis} + 2H\,\delta\Theta =-4\pi G\big(\delta\rho + 3\delta P  \big) + \cd^a A_a \,.
\end{align}

Expanding
$\mathbb Q=\mathbb X/\mathbb H^2-3$ to first order, we find that the $\ell>0$
 multipoles take the same form:
\begin{align}
\mathbb Q_{{\cal A}_\ell}
 =\frac{1}{H^2}\,\mathbb X_{{\cal A}_\ell}
 -\frac{2(q+3)}{H}\,\mathbb H_{{\cal A}_\ell}\,,
\end{align}
where $1 \le \ell \leq 3$ and  $\mathbb H_{abc}=0$   by \autoref{hub}.
The deceleration multipoles are then given to first order by}
\begin{align}
H^2\,  \langle\mathbb Q\rangle&=H^2\, q-\frac{2}{3} qH {\delta\Theta}
 +\frac{4\pi G}{3}\big( {\delta\rho} +3\,\delta P\big)-\frac23 \cd^aA_a\,,
 \label{Qmult1}\\
H^2\,  \mathbb Q_a&=\frac13\, \cd_a\Theta
+\frac25\,\cd^b \si_{ab} + (\dot A_a)_\pe+2(1+q)HA_a
  \,,
 \label{Qmult2a}\\
H^2\,  \mathbb Q_{ab}&=-{2q}{H}\si_{ab}+{E_{ab}}
 -4\pi G\, \pi_{ab}-2\,\cd_{\langle a}A_{b\rangle} \,,
 \label{Qmult3}\\
H^2\, \mathbb Q_{abc}&=\cd_{\langle a}\si_{bc\rangle}\,.
 \label{Qmult4}
\end{align}
Note that $\mathbb Q_{abcd}=0$ at linear order.
In the dipole \autoref{Qmult2a}, an alternative form for the first 2 terms on the right is given by the linearised version of \autoref{emce}: $\tfrac13\, \cd_a\Theta + \tfrac25\,\cd^b \si_{ab}= \tfrac35\cd_a\Theta +\tfrac25 {\rm curl}\, \om_a$. In the quadrupole we used the linearised version of
\autoref{emcf} to replace the time derivative  $(\dot\si_{ab})_\pe$ by algebraic terms.

Three features of \autoref{Qmult1}--\autoref{Qmult4} are worth highlighting.
\begin{itemize}
    \item
The
anisotropic pressure enters the quadrupole only in the combination
$E_{ab}-4\pi G\,\pi_{ab}$, so at this order it is  degenerate with the tidal field
and cannot be separated from it by a cosmographic   distance-redshift measurement.
\item
The nonlinear kinetic correction to the octupole includes the term $-H^{-2}A_{\langle a}\si_{bc\rangle}$ in \autoref{Xmult1}. Therefore the octupole differs from the dust form only at second order.

   \item
The dipole holds for any 4-acceleration. In the Vlasov case, the evolution of the 4-acceleration is determined self-consistently in the linearised  approximation:
\begin{align}\label{dota2x}
 (\dot A_a)_\pe + 3 H A_a= \frac{\bar P}{3(\bar\rho+\bar P)}\,\big(5\, \cd_a\Theta + 6\, \cd^b \si_{ab} \big) \,.
\end{align}
Note that we neglect terms proportional to $\bar P(\bar\rho+\bar P)^{-1} A_a$, which are of higher velocity order.
This evolution equation  allows us to eliminate $\dot A_a$, reducing \autoref{Qmult2a} to
\begin{align}
    H^2\,  \mathbb Q_a = (2q-1)H A_a + \Big[\frac{1}{15}+ \frac{\bar P}{3(\bar\rho+\bar P)}\Big]\, \big(5\, \cd_a\Theta + 6 \cd^b \si_{ab} \big)
  \,.
 \label{Qmult2b}
\end{align}
   \end{itemize}

~\\
\autoref{dota2x} is proved as follows.
First we linearise the 4-acceleration \autoref{da3} using \autoref{barr}:
\begin{align}\label{da3l}
  A_a=- \big(\bar \rho + \bar P\big)^{-1}
  \big(\cd_aP+\cd^b\pi_{ab}\big) \,.
\end{align}
Then we take  the covariant time derivative $u^b\nabla_b$  to obtain
\begin{align} \label{dota1}
  (\dot A_a)_\pe - \Big(3+ 5 \frac{\bar P}{\bar\rho+\bar P}  \Big)HA_a =
  -\frac{1}{\bar\rho+\bar P} \,\big[(\cd_a P)^{\dis}+ (\cd^b \pi_{ab})^{\dis} \big]_\pe\,.
\end{align}
For $\dot P$ and $(\dot\pi_{ab})_\pe$, we use the linearised \autoref{dp} and \autoref{dpi}. On the right-hand side, we require
the linearised covariant identities for commuting gradients and divergences with time derivatives \cite{Maartens:1998xg,Tsagas:2007yx}:
\begin{align} \label{covid}
    (\cd_a J)^{\dis}_\pe = \cd_a \dot J + A_a\,\dot{\bar J}  - H \cd_a J\,,\qquad
    (\cd^b J_{ab})^{\dis}_\pe =\cd^b \dot J_{ab} - H\, \cd^b J_{ab}\,.
\end{align}
Then \autoref{dota1} reduces to \autoref{dota2x},
using $\cd_a \dot P= - \tfrac53 \cd_a (\Theta P)=- \tfrac53 \bar P \cd_a\Theta -  5H  \cd_a P$ and $\cd^b \dot \pi_{ab}=-5H \cd^b \pi_{ab} -2\bar P \cd^b \si_{ab}$.

The evolution \autoref{dota2x} shows how the linearised 4-acceleration is sourced by isotropic stress at second velocity order, modulated by expansion gradients and shear divergence.

The vanishing of $\mathbb X_{abcd}$
enforced by \autoref{X}
deserves separate
emphasis because it is exact, requires neither field equations nor a matter
model, and survives both  4-acceleration and vorticity.  The hexadecapole of
$\mathbb Q$ inherits this result at linear perturbative order,
but not beyond: division by the direction-dependent
$\mathbb H^2$ generates composite higher multipoles even when the numerator
terminates.  A measured $\mathbb Q$ hexadecapole is therefore not by itself a
null test of general relativity; it must first be compared with these
nonlinear products.  Such a comparison can be made with the multipole
reconstructions of
\cite{Kalbouneh:2022tfw,Kalbouneh:2024szq,Kalbouneh:2025jnp}.

The Hubble multipoles give only the leading term in the local
distance--redshift relation.
Up to $O(z^2)$ for the estimator
\autoref{eq:main-eta-def}, we have \cite{Kalbouneh:2024yjj}
\begin{align}
 (\ln10)\, \eta(z,n^a)={}&\big(\ln\mathbb H
 -\left\langle\ln\mathbb H\right\rangle\big)_{\rm o}
 +\frac{z}{2}\big(\mathbb Q-\langle\mathbb Q\rangle\big)_{\rm o}
 \notag\\
 &+\frac{z^2}{24}\Big[
 4(\mathbb J-\mathbb R)-10\mathbb Q-9\mathbb Q^2
 -\left\langle4(\mathbb J-\mathbb R)-10\mathbb Q
 -9\mathbb Q^2\right\rangle\Big]_{\rm o}
 +O(z^3)\,.
 \label{eq:eta-higher-z}
\end{align}
Here $\mathbb J-\mathbb R$ is the covariant jerk--curvature combination \cite{Kalbouneh:2024yjj}.
Thus the $O(z)$ term is fixed by the deceleration multipoles, whereas the
$O(z^2)$ term contains genuinely new local information as well as products
of the lower-order coefficients.
Up to $O(z)$ the  4-acceleration contribution to the dipole is given by
\begin{align}
H (\ln10)\, \eta_A(z,n^a)\big|_{\rm dipole} &= -\big(n^a A_a\big)_{\rm o}
 \\ \notag
 &~~~
 +\frac{z}{2}\,\bigg[n^a
 \bigg\{(2q-1) A_a +\frac{\bar P}{3H(\bar\rho+\bar P)}\, \big(5\cd_a\Theta + 6 \cd^b\si_{ab} \big)  \bigg\}
 \bigg]_{\rm o}   +O(z^2)\,,
\end{align}
{where we used \autoref{dota2x} to eliminate the time derivative of $A_a$.}
The acceleration contributions to the monopole \autoref{Qmult1} and quadrupole \autoref{Qmult3} are proportional to
\begin{align}
    \cd^aA_a&= -(\bar\rho + \bar P)^{-1}\big(\cd^2 P+ \cd^a \cd^b \pi_{ab} \big)\,,
    \\
    \cd_{\la a} A_{b\ra}&= -(\bar\rho + \bar P)^{-1}\big( \cd_{\la a} \cd_{b\ra} P +
    \cd_{\la a} \cd^c \pi_{b\ra c}\big)\,.
\end{align}

\newpage
\section{Conclusion}
\label{sec:conclusion}
We have constructed a covariant kinetic extension of local dust
cosmography.  The exact Vlasov equation generates a two-dimensional hierarchy
of energy-integrated velocity moments.  Organising this hierarchy by velocity
weight makes the approximation transparent: up to $O(v^2\rho)$ the matter
energy frame closes on $\rho$, $P$ and $\pi_{ab}$, whereas the third central
moment is the first omitted dynamical sector.  This is a controlled statement
about the matter model and does not require a background geometry or a
perturbative expansion of the spacetime.

Within this closure, the momentum equation fixes a nonzero matter-frame
4-acceleration from gradients of the isotropic and anisotropic kinetic
stresses and we provide an exact nonlinear expression for the 4-acceleration.  Acceleration in turn sources vorticity only through nonlinear
scalar couplings or vector stress components.  In local cosmography it appears
directly as the intrinsic dipole of the covariant Hubble parameter.  We have
also derived its contributions to the covariant deceleration multipoles and
shown how the Hubble multipoles transform to a boosted observer.  These results
distinguish the physical non-geodesicity of the collisionless matter frame from
the kinematic dipole produced by observer motion.

The novelty is therefore not the general possibility of a non-geodesic
congruence, but its derivation from a specified collisionless matter model and
its consistent propagation through the local cosmographic hierarchy.  The
framework supplies the dynamical completion needed to interpret a measured
intrinsic Hubble dipole as a stress effect rather than merely a choice of
frame.
{In addition, our analysis provides the basis for testing whether the breakdown in fitting Cosmicflows-4 data within covariant dust cosmography   at $z\lesssim0.01$ \cite{Kalbouneh:2025jnp}, can be cured by velocity dispersion.}

The immediate limitations of this approach are equally clear: the second-moment closure
must fail when higher velocity cumulants become important, and a galaxy sample
need not trace the matter energy frame exactly.  The companion perturbative
paper tests the former limitation, predicts the window-dependent acceleration
and the induced dipole, quadrupole and octupole corrections, and identifies
tracer response as the remaining step towards a catalogue-level comparison.

\vspace*{0.5cm}
\acknowledgments
CC is supported by STFC (UK) grant ST/X000931/1.\\
During the development of this work, the authors used OpenAI Codex (GPT-5.6 Sol and 6.0 Astra) and Anthropic Claude (Claude Opus 5) to assist with algebraic consistency checks. CC thanks OpenAI for providing complimentary access to ChatGPT through the ChatGPT for Academic Researchers programme.

\clearpage
\providecommand{\href}[2]{#2}\begingroup\raggedright\endgroup
\end{document}